# *In silico* identification of potential natural product inhibitors of human proteases key to SARS-CoV-2 infection

R.P. Vivek-Ananth[1,2], Abhijit Rana[2,3], Nithin Rajan[1], Himansu S. Biswal[2,3,*], Areejit Samal[1,2,4,*,$]

*[1]The Institute of Mathematical Sciences (IMSc), Chennai 600113, India*

*[2]Homi Bhabha National Institute (HBNI), Mumbai 400094, India*

*[3]School of Chemical Sciences, National Institute of Science Education and Research (NISER), Bhubaneswar 752050, India*

*[4]Max Planck Institute for Mathematics in the Sciences, Leipzig 04103, Germany*

*[*]Correspondence to: asamal@imsc.res.in or himansu@niser.ac.in*

*[$]Lead contact*

**Address of Lead Contact**

Areejit Samal

The Institute of Mathematical Sciences (IMSc),

4[th] Cross Road, CIT Campus, Taramani,

Chennai 600113, India

Phone: +91-44-22543219

Email: asamal@imsc.res.in

URL: https://www.imsc.res.in/~asamal/

# Abstract

Presently, there are no approved drugs or vaccines to treat COVID-19 which has spread to over 200 countries and is responsible for over 3,65,000 deaths worldwide. Recent studies have shown that two human proteases, TMPRSS2 and cathepsin L, play a key role in host cell entry of SARS-CoV-2. Importantly, inhibitors of these proteases were shown to block SARS-CoV-2 infection. Here, we perform virtual screening of 14010 phytochemicals produced by Indian medicinal plants to identify natural product inhibitors of TMPRSS2 and cathepsin L. We built a homology model of TMPRSS2 as an experimentally determined structure is not available. AutoDock Vina was used to perform molecular docking of phytochemicals against TMPRSS2 model structure and cathepsin L crystal structure. Potential phytochemical inhibitors were filtered by comparing their docked binding energies with those of known inhibitors of TMPRSS2 and cathepsin L. Further, the ligand binding site residues and non-covalent protein-ligand interactions were used as an additional filter to identify phytochemical inhibitors that either bind to or form interactions with residues important for the specificity of the target proteases. We have identified 96 inhibitors of TMPRSS2 and 9 inhibitors of cathepsin L among phytochemicals of Indian medicinal plants. The top inhibitors of TMPRSS2 are Edgeworoside C, Adlumidine and Qingdainone, and of cathepsin L is Ararobinol. Interestingly, several herbal sources of identified phytochemical inhibitors have antiviral or anti-inflammatory use in traditional medicine. Further *in vitro* and *in vivo* testing is needed before clinical trials of the promising phytochemical inhibitors identified here.

## 1. Introduction

In December 2019, a new respiratory disease with unknown cause was first reported in Wuhan, China with clinical symptoms of fever, cough, shortness of breath, fatigue and pneumonia (Huang et al., 2020; C. Wang et al., 2020; Zhu et al., 2020). While most cases of this new disease show mild to moderate symptoms, a small fraction of cases, especially those with comorbid conditions like diabetes and hypertension, can develop fatal conditions such as acute respiratory distress syndrome (ARDS) due to severe lung damage (Chan et al., 2020). In January 2020, a novel betacoronavirus, initially named 2019-nCoV, was discovered to be the etiological agent of this new disease (Huang et al., 2020; C. Wang et al., 2020; Zhu et al., 2020). Subsequently, human-to-human transmission of this disease was confirmed (Chan et al., 2020). By 30 January 2020, the 2019-nCoV had spread to more than 20 countries and the world health organization (WHO) declared a public health emergency of international concern. On 11 February 2020, the international committee on taxonomy of viruses permanently named 2019-nCoV as severe acute respiratory syndrome coronavirus 2 (SARS-CoV-2) and the WHO named the associated disease as coronavirus disease 2019 (COVID-19). By 11 March 2020, COVID-19 had spread to more than 150 countries across 6 continents and the WHO upgraded the status of the epidemic to pandemic. As of 30 May 2020, the number of laboratory confirmed COVID-19 cases and deaths have already surpassed 6 million and 3,65,000, respectively, with the worst affected countries as USA, UK, Italy, France, Spain and Brazil (https://www.worldometers.info/coronavirus/). In short, the COVID-19 pandemic poses an unprecedented public health and economic threat to humankind.

Coronaviruses are enveloped, positive-sense, single-stranded RNA viruses with large viral genomes. The publication of SARS-CoV-2 genome in January 2020 led to its taxonomic classification into the family *Coronaviridae* and genus *Betacoronavirus* (Huang et al., 2020; C. Wang et al., 2020; Zhu et al., 2020). Bats are natural reservoirs of coronaviruses (W. Li et al., 2005). The SARS-CoV-2 genome shares 96% nucleotide identity with bat coronavirus RaTG13, which suggests a probable zoonotic transfer to humans via an intermediate animal host (Zhou et al., 2020). Prior to SARS-CoV-2 epidemic, there are two precedences of zoonotic transfer of betacoronaviruses to humans, namely the severe acute respiratory syndrome coronavirus (SARS-CoV) and the middle east respiratory syndrome coronavirus (MERS-CoV), which had also led to outbreaks of severe respiratory disease (de Wit et al., 2016). In 2002-2003, SARS-CoV emerged in China and led to 8098 infections and 774 deaths across the world (de Wit et al., 2016). Interestingly, the SARS-CoV-2 genome shares ~80% nucleotide identity with SARS-CoV (Huang et al., 2020; C. Wang et al., 2020; Zhu et al., 2020). In 2012, MERS-CoV emerged in Saudi Arabia and led to 2521 infections and 866 deaths that were largely limited to middle eastern countries (de Wit et al., 2016). Unlike SARS and MERS, the geographic spread of COVID-19 and the ensuing mortality is significantly higher. To date, there are no approved antiviral drugs or vaccines against betacoronavirus infections including COVID-19 (G. Li & De Clercq, 2020). Hence, the current measures to contain COVID-19 include social distancing, aggressive testing, patient isolation, contact tracing and travel restrictions. In present circumstances, an immediate goal of biomedical research is to develop antivirals or anti-COVID therapeutics for SARS-CoV-2 (G. Li & De Clercq, 2020).

The SARS-CoV-2 genome comprises ~30000 bases with 14 open reading frames (ORFs) coding for 27 proteins (A. Wu et al., 2020). The genome organization of SARS-CoV-2 is similar to other betacoronaviruses with the 5'-region coding for non-structural proteins and the 3'-region coding for structural proteins. Important structural proteins of SARS-CoV-2 coded by the 3'-region include the spike (S) surface glycoprotein, the envelope (E) protein, the matrix (M) protein and the nucleocapsid (N) protein (A. Wu et al., 2020). The 5'-region of the SARS-CoV-2 genome contains the replicase gene which codes for two overlapping polyproteins, pp1a and pp1ab, which are proteolytically cleaved by two important non-structural proteins, 3-chymotrypsin like protease (3CL$^{pro}$) and papain-like protease (PL$^{pro}$), to produce functional (non-structural) proteins. Other important non-structural proteins of SARS-CoV-2 for the viral life cycle include the RNA-dependent RNA polymerase (RdRp) and helicase. Accordingly, the 4 non-structural proteins, 3CL$^{pro}$, PL$^{pro}$, RdRp and helicase, along with the spike glycoprotein of SARS-CoV-2 are among the most attractive targets for anti-COVID drugs (G. Li & De Clercq, 2020).

Rather than targeting important SARS-CoV-2 proteins for viral life cycle, an alternative approach to anti-COVID drugs involves targeting host factors key to SARS-CoV-2 infection (G. Li & De Clercq, 2020). For host cell entry, SARS-CoV-2 employs the spike (S) protein whose S1 subunit has a receptor binding domain (RBD) that specifically recognizes the cell surface receptor angiotensin converting enzyme 2 (ACE2) (Hoffmann, Kleine-Weber, et al., 2020; Ou et al., 2020; Shang, Wan, et al., 2020; Shang, Ye, et al., 2020; Yan et al., 2020). Notably, both SARS-CoV-2 and SARS-CoV employ ACE2 as the cell entry receptor (Hoffmann,

Kleine-Weber, et al., 2020; Ou et al., 2020; Shang, Wan, et al., 2020; Shang, Ye, et al., 2020; Yan et al., 2020). After attachment of S protein to ACE2 receptor, the membrane fusion of virus and host cell depends on proteolytic activation of S protein by host proteases which involves cleavage of S1 subunit at S1/S2 and S2' sites resulting in dissociation of S1 subunit and structural change in S2 subunit of S protein (Hoffmann, Kleine-Weber, et al., 2020; Ou et al., 2020; Shang, Wan, et al., 2020; Shang, Ye, et al., 2020; Yan et al., 2020). Hoffmann *et al* (Hoffmann, Kleine-Weber, et al., 2020) showed that the host cell proteases, Transmembrane Protease Serine 2 (TMPRSS2) and cathepsin B or cathepsin L, can carry out S protein priming required for SARS-CoV-2 entry. Hoffmann *et al* (Hoffmann, Kleine-Weber, et al., 2020) also showed that TMPRSS2 is more essential for S protein priming and SARS-CoV-2 entry. In parallel, Ou *et al* (Ou et al., 2020) used specific inhibitors of cathepsin L and cathepsin B to show that cathepsin L rather than cathepsin B is essential for S protein priming of SARS-CoV-2 and membrane fusion in lysosomes. These studies highlight at least two alternate pathways for host cell entry of SARS-CoV-2. On the one hand, after SARS-CoV-2 attachment to ACE2, the membrane fusion and cytoplasmic entry can occur at the plasma membrane provided the cell surface protease TMPRSS2 is available to carry out S protein priming (Hoffmann, Kleine-Weber, et al., 2020; Ou et al., 2020; Shang, Wan, et al., 2020). On the other hand, after SARS-CoV-2 attachment to ACE2, the virus can be internalized as part of endosomes in the endocytic pathway, and later, the membrane fusion and cytoplasmic entry will occur in lysosomes provided the lysosomal protease cathepsin L is available to carry out S protein priming (Hoffmann, Kleine-Weber, et al., 2020; Ou et al., 2020; Shang, Wan, et al., 2020). Depending on the target cell and associated expression of host cell proteases, SARS-CoV-2 may use one of the alternative pathways for host cell entry (Hoffmann, Kleine-Weber, et al., 2020; Ou et al., 2020; Shang, Wan, et al., 2020). Importantly, the above-mentioned studies also showed that known inhibitors of TMPRSS2 and cathepsin L can block or significantly reduce the host cell entry of SARS-CoV-2 (Hoffmann, Kleine-Weber, et al., 2020; Ou et al., 2020; Shang, Wan, et al., 2020). In conclusion, human proteases TMPRSS2 and cathepsin L are key factors for host cell entry and are important targets for anti-COVID drugs (Hoffmann, Kleine-Weber, et al., 2020; Ou et al., 2020; Shang, Wan, et al., 2020; Hoffmann, Schroeder, et al., 2020).

To expedite this search for anti-COVID drugs, several computational studies have used homology modeling or published crystal structures of SARS-CoV-2 proteins, molecular docking and diverse small molecule libraries, to predict potential inhibitors of SARS-CoV-2 proteins including among existing approved drugs for repurposing and natural compounds (see e.g. (Elfiky, 2020; Islam et al., 2020; Shah et al., 2020; C. Wu et al., 2020)). In comparison, fewer computational studies (Rahman et al., 2020; C. Wu et al., 2020) have focussed on identification of potential inhibitors of host factors. In this work, we perform virtual screening of a large phytochemical library specific to Indian medicinal plants to identify potential natural product inhibitors of TMPRSS2 and cathepsin L.

Plant-based natural products have made immense contributions to drug discovery (Newman & Cragg, 2012). Specifically, ~ 40% of the small-molecule drugs approved to date by the US Food and Drug Administration (FDA) are either natural products or natural product derivatives. Recently, there are reports from China of successful use of traditional Chinese medicine and associated herbs in treatment of COVID-19 patients (Ren et al., 2020). On similar lines, there have been suggestions to tap the rich legacy of traditional Indian medicine and information on phytochemicals of Indian herbs in the search for anti-COVID drugs (Vellingiri et al., 2020). Previously, some of us have built IMPPAT, the largest resource to date on phytochemicals of Indian herbs (Mohanraj et al., 2018). In this work, we perform molecular docking using a large library of 14010 phytochemicals compiled mainly from IMPPAT to identify potential natural product inhibitors of TMPRSS2 and cathepsin L.

## 2. Materials and Methods

### 2.1 Phytochemical library and drug-likeness evaluation

Previously, some of us have built the Indian Medicinal Plants, Phytochemicals And Therapeutics (IMPPAT) database (Mohanraj et al., 2018) which is the largest resource on phytochemicals of Indian herbs to date. For this study, we compiled a ligand library of 14011 phytochemicals by augmenting the information in IMPPAT (Mohanraj et al., 2018) with additional information compiled from other literature sources (Rastogi, 1990, 1991, 1993, 1995, 1998; *The Wealth of India*, 2000a; *The Wealth of India*, 2000b; *The Wealth of India*, 2000c; *The Wealth of India*, 2000d; *The Wealth of India*, 2000e). Thereafter, the widely-used drug-likeness measure, Lipinski's rule of five (RO5) (Lipinski et al., 2001), was employed to filter the potential drug-like molecules within the ligand library of 14011 phytochemicals. Specifically, 10510 phytochemicals passed the R05 drug-likeness filter. We then retrieved the three-dimensional (3D) structures of these phytochemicals from Pubchem (Kim et al., 2019). Next the 3D structures of the drug-like phytochemicals were energy-minimized using *obminimize* within the OpenBabel toolbox (O'Boyle et al., 2011). Finally, the energy-minimized 3D structures of ligands in .sdf format were converted to .pdb format using OpenBabel.

## 2.2 Homology modeling of TMPRSS2 structure

TMPRSS2 is a trypsin-like serine protease whose catalytic site consists of the triad Ser441 (S441), His296 (H296) and Asp345 (D345) (Paoloni-Giacobino et al., 1997). It is well established that trypsin-like serine proteases cleave peptide bonds following positively charged amino acid residues such as arginine or lysine, and this specificity of the enzyme is determined by a negatively charged aspartate residue located at the bottom of its S1 pocket (Evnin et al., 1990). In TMPRSS2, this specificity is determined by the conserved negatively charged residue Asp435 (D435) at the bottom of the S1 pocket (Paoloni-Giacobino et al., 1997).

To date the 3D structure of TMPRSS2 has not been experimentally determined, and thus, we have used SWISS-MODEL (Waterhouse et al., 2018) webserver (https://swissmodel.expasy.org/interactive) to perform homology modeling of TMPRSS2. We submitted the TMPRSS2 protein sequence (NCBI reference sequence NP_005647.3) to SWISS-MODEL and selected the crystal structure of human protein hepsin (PDB 1Z8G) (Herter et al., 2005) as the template to build the model structure (Figure 1a). Note that hepsin is also a Type II transmembrane trypsin-like serine protease, and it shares 38% sequence similarity with TMPRSS2 (Figure 1b). Subsequently, UCSF Chimera (Pettersen et al., 2004) was used to minimize the energy of the TMPRSS2 model structure obtained from SWISS-MODEL. Thereafter, the energy-minimized TMPRSS2 model structure was assessed using the structure assessment tool within SWISS-MODEL. In the TMPRSS2 model, 94.19% of the amino acid residues were found to be in the Ramachandran favoured regions in the Ramachandran plot (Figure 1c) and the model structure has a MolProbity (Chen et al., 2010) score of 1.50.

## 2.3 Protein structure of cathepsin L

We use the crystal structure (PDB 5MQY) (Kuhn et al., 2017) of human cathepsin L with 1.13 Å resolution obtained from Protein Data Bank for virtual screening. UCSF Chimera was used to minimize the energy of the cathepsin L structure. Figure 2 displays the cathepsin L structure with important residues in S1, S2 and S1' subsites of the enzyme (Fujishima et al., 1997). Previous research has also revealed that S1 and S2 subsites of cathepsin L are important for the specificity of the enzyme (Fujishima et al., 1997; Vito Turk et al., 2012). In cathepsin L, the catalytic site consists of Cys25 (C25) and His163 (H163) in the S1 subsite, and Trp189 (W189) is at the center of the S1' subsite (Fujishima et al., 1997) (Figure 2). In cathepsin L, the S2 subsite with important residues Asp162 (D162), Met161 (M161), Ala135 (A135), Met70 (M70) and Leu69 (L69) forms a deep hydrophobic pocket, and lastly, the conserved residue Gly68 (G68) is at the center of the S3 subsite (Adams-Cioaba et al., 2011; Fujishima et al., 1997) (Figure 2).

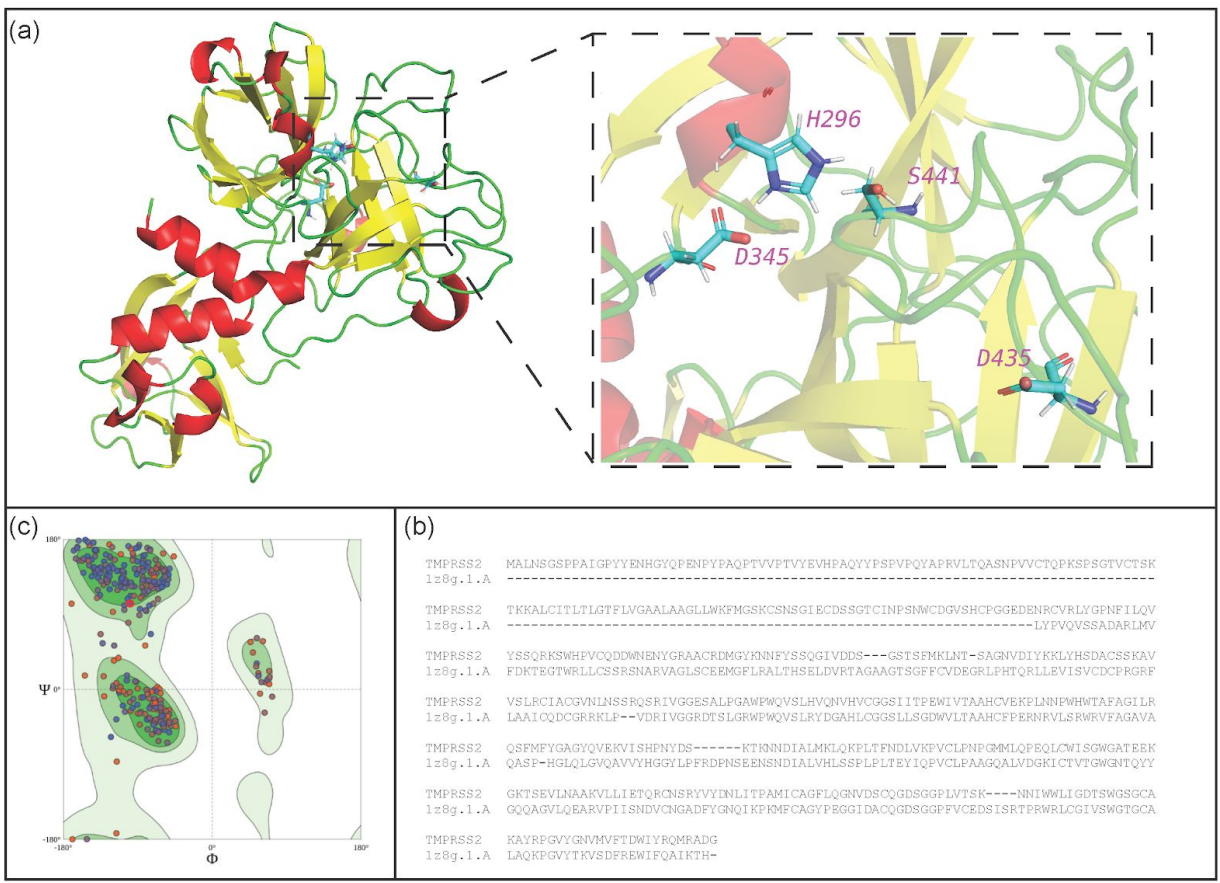

**Figure 1. (Colour online)** (a) Cartoon representation of the homology model structure of TMPRSS2 which has been energy-minimized using UCSF Chimera. The figure zooms into the region containing the catalytic triad Ser441 (S441), His296 (H296) and Asp345 (D345), and the substrate binding residue Asp435 (D435) in the S1 subsite of the enzyme. (b) Alignment of protein sequences for TMPRSS2 and hepsin (PDB 1Z8G) which was used as a template to model the structure of TMPRSS2. (c) General Ramachandran plot of the energy-minimized model structure of TMPRSS2, which displays the torsional angles, phi (φ) and psi (ψ), of the amino acid residues in the protein.

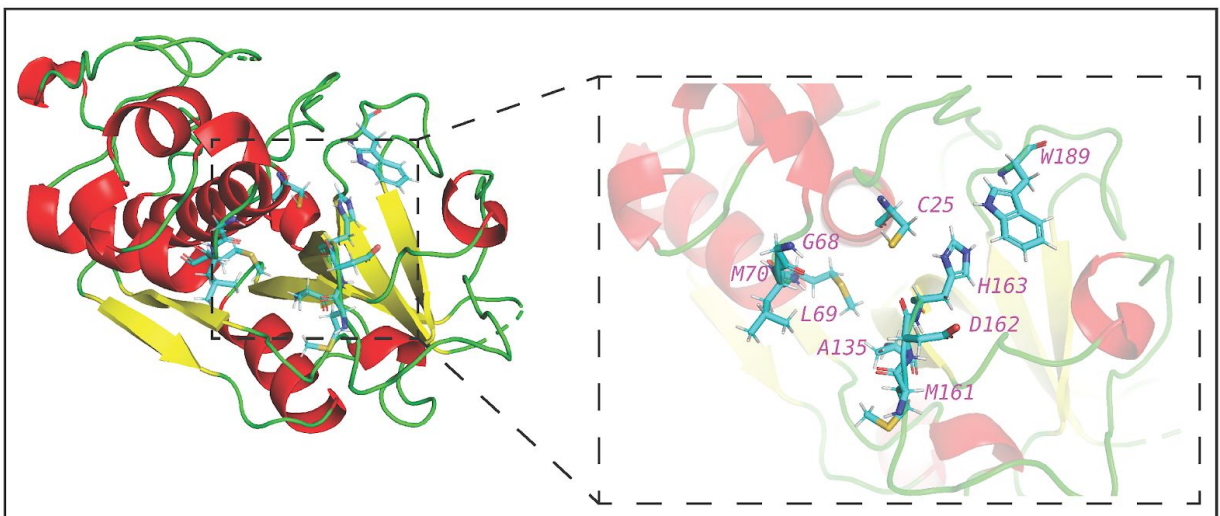

**Figure 2. (Colour online)** Cartoon representation of the crystal structure of human cathepsin L (PDB 5MQY). The figure zooms into the region containing the catalytic residues Cys (C25) and His163 (H163) in the S1 subsite, residues Asp162 (D162), Met161 (M161), Ala135 (A135), Met70 (M70) and Leu (L69) in the S2 subsite, Trp189 (W189) at the centre of S1' subsite and the conserved residue Gly68 (G68) in the S3 subsite of the enzyme.

## 2.4 Molecular Docking

AutoDock Vina (Trott & Olson, 2010) was used to perform the molecular docking of energy-minimized 3D structures of ligands with energy-minimized structure of target proteins. Accordingly, the 3D structures of prepared ligands in .pdb format were converted to .pdbqt format using the python script *prepare_ligand4.py* from AutoDockTools (Morris et al., 2009). Similarly, the energy-minimized structure of TMPRSS2 and cathepsin L in .pdb format were converted to .pdbqt format using the python script *prepare_receptor4.py* from AutoDockTools (Morris et al., 2009).

For protein-ligand docking, the appropriate grid box specified by the search space centre and search space size for TMPRSS2 and cathepsin L was manually determined by considering the key residues in target proteins such as the catalytic residues and substrate binding residues, which are important for the function and specificity of the considered proteases as reported in the literature. For TMPRSS2, the grid box was defined by the search space centre (40.50, -6.00, 25.70) and search space size of 25Å x 25Å x 25Å. For cathepsin L, the grid box was defined by the search space centre (55.06, 48.65, 18.12) and search space size of 25Å x 25Å x 25Å.

For both target proteins, the molecular docking of prepared ligands was performed using AutoDock Vina with the exhaustiveness set as 8. The top conformation of the docked ligand with lowest binding energy, i.e. the best docked pose, was obtained from the output of AutoDock Vina using the associated script *vina_split*. Subsequently, a combined structure file in .pdb format of the docked protein-ligand complex (with ligand in the best docked pose) was prepared using a custom script and pdb-tools (Rodrigues et al., 2018).

## 2.5 Identification of Protein-Ligand interactions

We searched the combined structure file of the docked protein-ligand complex for ligand binding residues in protein and different non-covalent interactions that can facilitate the binding of the ligand with the protein. These non-covalent protein-ligand interactions were identified using different geometric criteria which are specific to different types of interactions.

*Binding site residue*. Ligand binding site residues are defined as amino acids in protein which have at least one non-hydrogen atom in the proximity of at least one non-hydrogen atom of the ligand. The distance cut off to determine this proximity between non-hydrogen atoms of protein and ligand is taken to be the sum of their van der Waals radius plus 0.5Å (Schmidt et al., 2011).

*Hydrogen bonds*. The accepted geometric criteria for hydrogen bonds of type D-H···A are as follows. Firstly, the distance between the hydrogen (H) and acceptor (A) atom should be less than the sum of their van der Waals radii. Secondly, the angle formed by donor (D), H and A atoms should be > 90° (Supplementary Figure S1a). Moreover, carbon (C), nitrogen (N), oxygen (O) or sulfur (S) atoms can be donors while N, O or S atoms can be acceptors (Sarkhel & Desiraju, 2004; Zhou et al., 2009).

*Chalcogen bonds*. In contrast to hydrogen bonds, chalcogen bonds are of type C-Y···A, where Y can be a S or selenium (Se) atom and A can be a N, O or S atom. The accepted geometric criteria for chalcogen interactions are as follows. Firstly, the distance between Y and A should be less than the sum of their van der Waals radii. Secondly, the angle formed by the triad, that is ∠C-Y···A, should lie in the range 150° to 180° (Supplementary Figure S1b) (Kříž et al., 2018).

*Halogen bonds*. Halogen bonds are of type C-Y···A-B, where halogen Y can be a Fluorine (F), Chlorine (Cl), Bromine (Br) or Iodine (I) atom and A can be a N, O or S atom. The formation of the halogen bond is favoured when the distance between Y and A is $\leq 3.7$ Å and the angle $\theta_1$ of the A atom relative to the C-Y bond, and the angle $\theta_2$ of the halogen Y relative to the A-B bond should be $\geq 90°$ (Supplementary Figure S1c) (Borozan & Stojanović, 2013).

*π-π stacking*. This interaction occurs between two aromatic rings and can be majorly classified into two types, namely, face-to-face and face-to-edge. In the case of face-to-face type of *π-π* interaction, the distance between the centroids of the two participating aromatic rings should be < 4.4 Å and the angle between their ring planes should be < 30°. In the case of face-to-edge type of *π-π* interaction, the distance between the centroids of the two participating aromatic rings should be < 5.5 Å and the angle formed by the ring planes should be in the range 60° to 120° (Supplementary Figure S1d,e).

*Hydrophobic interactions*. The geometric criteria for the formation of hydrophobic interactions between atoms in protein and ligand are as follows (Ferreira de Freitas & Schapira, 2017). The distance between a carbon atom in protein or ligand and a carbon, halogen or sulfur atom in ligand or protein, respectively, should be $\leq 4$ Å. Furthermore, we ensure that the involved atoms in a hydrophobic interaction between protein and ligand do not form hydrogen, chalcogen or halogen bonds between them (Ferreira de Freitas & Schapira, 2017).

In order to detect the above-mentioned protein-ligand interactions, an in-house Python program was written to enable batch processing of combined structure files containing docked protein-ligand complexes for our large phytochemical library.

## 2.6  Comparison with reference inhibitors of TMPRSS2 and cathepsin L

In order to identify potent phytochemical inhibitors of target proteins, we decided to compare the binding energy of the best docked pose of ligands with binding energies of the best docked pose of known inhibitors of TMPRSS2 and cathepsin L obtained from AutoDock Vina.

Recent experiments have shown that both camostat mesylate and nafamostat mesylate, which are approved for human use in Japan, can block the TMPRSS2-dependent cell entry of SARS-CoV-2 (Hoffmann, Kleine-Weber, et al., 2020; Hoffmann, Schroeder, et al., 2020). By docking these two inhibitors to TMPRSS2 using AutoDock Vina with exhaustiveness set at 50, the predicted binding energies of camostat and nafamostat was found to be -7.4 kcal/mol and -8.5 kcal/mol, respectively. Supplementary Figures S2a,b show the best docked poses of nafamostat and camostat with TMPRSS2, and it is seen that both molecules form hydrogen bonds with the substrate binding residue D435. Importantly, in comparison to camostat mesylate, nafamostat mesylate in a recent experiment was shown to inhibit the TMPRSS2-dependent cell entry with 15-fold higher efficiency and an $EC_{50}$ value in lower nanomolar range (Hoffmann, Schroeder, et al., 2020), and thus, the docked binding energies of these two known inhibitors are in line with experiments. Based on above observations, we decided on a stringent criteria of docked binding energy $\leq$ -8.5 kcal/mol for screened ligands to be identified as potential inhibitors of TMPRSS2.

Recent experiments have shown that the small molecules E-64d and PC-0626568 (SID26681509) can block the cathepsin L-dependent cell entry of SARS-CoV-2 (Hoffmann, Kleine-Weber, et al., 2020; Ou et al., 2020; Shang, Wan, et al., 2020). Note that cathepsin L is one of 11 cysteine cathepsin proteases encoded by the human genome, and the cathepsins share a high sequence similarity to papain, a non-specific plant protease (V. Turk et al., 2001). E64-d is a broad spectrum inhibitor which can inhibit proteases cathepsins B, H, L and calpain, while PC-0626568 is a specific inhibitor of cathepsin L (Ou et al., 2020). Moreover, a recent study (Ou et al., 2020) used the specific inhibitor PC-0626568 of cathepsin L to conclude that cathepsin L rather than cathepsin B is important for cell entry of SARS-CoV-2. As both cathepsin L and cathepsin B are expressed in several mammalian tissues, it is important to design specific inhibitors of cathepsin L (Fujishima et al., 1997; Vito Turk et al., 2012) to avoid any off target toxicity. By docking these two known inhibitors to cathepsin L using AutoDock Vina with exhaustiveness set at 50, the predicted binding energies of E-64d and PC-0626568 was found to be -5.0 kcal/mol and -8.0 kcal/mol, respectively, and thus, the docked binding energies are in line with known specificity of E-64d and PC-0626568 to cathepsin L [11]. Supplementary Figures S2c,d show the best docked poses of PC-0626568 and E-64d with cathepsin L. It is seen that E-64d forms hydrogen bonds with both catalytic residues C25 and H163, whereas PC-0626568 forms a hydrogen bond with H163 and hydrophobic interaction with C25. Based on above observations, we decided on a stringent criteria of docked binding energy ≤ -8.0 kcal/mol for screened ligands to be identified as potential inhibitors of cathepsin L.

**2.7 Prediction of ADMET properties**

In order to assess the pharmacokinetic properties and potential toxicity of the inhibitors of TMPRSS2 and cathepsin L predicted from this docking study, we have also computed the Absorption, Distribution, Metabolism, Excretion and Toxicity (ADMET) properties of the inhibitors using SwissADME (Daina et al., 2017) and vNN-ADMET (Schyman et al., 2017).

**3. Results and Discussion**

**3.1 Workflow for virtual screening**

This computational study aims to predict potential phytochemical inhibitors of human proteases, TMPRSS2 and cathepsin L, that are important for priming of S protein and cell entry of SARS-CoV-2 (Hoffmann, Kleine-Weber, et al., 2020; Ou et al., 2020; Shang, Wan, et al., 2020). Briefly, the workflow for this virtual screening is as follows (Figure 3).

In the first stage, we prepared the ligands for molecular docking with target proteases. We compiled a library of 14011 phytochemicals produced by medicinal plants used in traditional Indian medicine, and the main source of this compilation was IMPPAT (Mohanraj et al., 2018), the largest resource on phytochemicals of Indian herbs to date (Methods). Next, the standard drug-likeness measure, Lipinski's rule of five (RO5) (Lipinski et al., 2001), was used to filter a subset of 10510 drug-like phytochemicals. Next, the filtered phytochemicals were prepared for docking by retrieving their 3D structures from Pubchem followed by energy-minimization using OpenBabel (O'Boyle et al., 2011) (Methods).

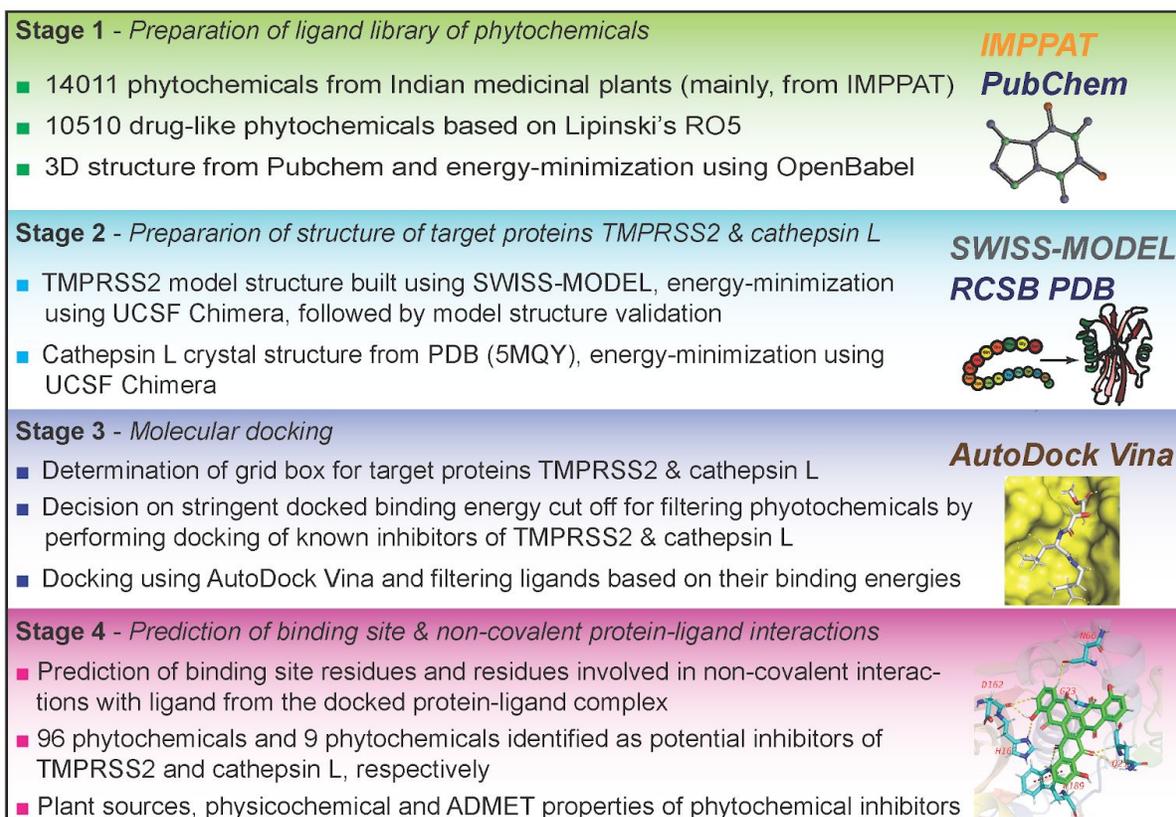

**Stage 1** - *Preparation of ligand library of phytochemicals*
- 14011 phytochemicals from Indian medicinal plants (mainly, from IMPPAT)
- 10510 drug-like phytochemicals based on Lipinski's RO5
- 3D structure from Pubchem and energy-minimization using OpenBabel

*IMPPAT*
**PubChem**

**Stage 2** - *Prepararion of structure of target proteins TMPRSS2 & cathepsin L*
- TMPRSS2 model structure built using SWISS-MODEL, energy-minimization using UCSF Chimera, followed by model structure validation
- Cathepsin L crystal structure from PDB (5MQY), energy-minimization using UCSF Chimera

*SWISS-MODEL*
**RCSB PDB**

**Stage 3** - *Molecular docking*
- Determination of grid box for target proteins TMPRSS2 & cathepsin L
- Decision on stringent docked binding energy cut off for filtering phytochemicals by performing docking of known inhibitors of TMPRSS2 & cathepsin L
- Docking using AutoDock Vina and filtering ligands based on their binding energies

*AutoDock Vina*

**Stage 4** - *Prediction of binding site & non-covalent protein-ligand interactions*
- Prediction of binding site residues and residues involved in non-covalent interactions with ligand from the docked protein-ligand complex
- 96 phytochemicals and 9 phytochemicals identified as potential inhibitors of TMPRSS2 and cathepsin L, respectively
- Plant sources, physicochemical and ADMET properties of phytochemical inhibitors

Figure 3

**Figure 3. (Colour online)** Virtual screening workflow to identify potential phytochemical inhibitors of human proteases TMPRSS2 and cathepsin L.

In the second stage, we prepared the target proteins for docking with prepared ligands. For TMPRSS2, the 3D structure is yet to be determined experimentally, and thus, we built a homology model of TMPRSS2 using SWISS-MODEL (Waterhouse et al., 2018) which was used for docking after energy-minimization using UCSF Chimera (Pettersen et al., 2004) (Figure 1; Methods). Figure 1 displays the TMPRSS2 model structure with the catalytic triad S441, H296 and D345 and the substrate binding residue D435 in the S1 subsite. For cathepsin L, the crystal structure (PDB 5MQY) [15] with 1.13 Å resolution was used for docking after energy-minimization using UCSF Chimera (Figure 2; Methods). Figure 2 displays the cathepsin L structure with the catalytic dyad C25 and H163 in S1 subsite, and other important residues in S2 and S1' subsites.

In the third stage, we performed protein-ligand docking using AutoDock Vina (Trott & Olson, 2010). For protein-ligand docking, an appropriate grid box was manually determined for TMPRSS2 and cathepsin L (Methods). To decide on a stringent binding energy cut off for the identification of potential inhibitors, docking was first performed for known inhibitors of target proteins (Methods). Based on the docking binding energies of the known inhibitors, camostat and nafamostat, to TMPRSS2, we decided on a stringent criteria of binding energy $\leq$ -8.5 kcal/mol for the best docked pose of screened ligands to identify potential inhibitors of TMPRSS2 (Methods). Similarly, based on the docking binding energies of the known inhibitors, E-64d and PC-0626568, to cathepsin L, we decided on a stringent criteria of binding energy $\leq$ -8.0 kcal/mol for the best docked pose of screened ligands to identify potential inhibitors of cathepsin L (Methods). Thereafter, docking was performed for the prepared ligands in the phytochemical library against the prepared structures of TMPRSS2 and cathepsin L (Methods). Lastly, we filtered the subset of phytochemicals whose binding energy in the best docked pose with TMPRSS2 (respectively, cathepsin L) is $\leq$ -8.5 kcal/mol (respectively, $\leq$ -8.0 kcal/mol). Moreover, the best docked pose with TMPRSS2 or cathepsin L of each filtered phytochemical was separated from AutoDock Vina output file, and then, combined with the target protein structure to obtain the docked protein-ligand complex in .pdb format (Methods). At the end of third stage, we obtained 101 phytochemicals whose binding energy in the best docked pose with TMPRSS2 is $\leq$ -8.5 kcal/mol and 16 phytochemicals whose binding energy in the best docked pose with cathepsin L is $\leq$ -8.0 kcal/mol.

In the fourth stage, the structure of docked protein-ligand complex in .pdb format for each filtered phytochemical from third stage was used to determine the ligand binding site residues in the target protein and different non-covalent interactions between ligand and target protein (Methods). In case of TMPRSS2, the specificity of this trypsin-like protease is determined by the conserved substrate binding residue D435 in the S1 pocket (Paoloni-Giacobino et al., 1997) (Methods), and therefore, a potent inhibitor should either bind or form non-covalent interactions with D435. In case of cathepsin L, the specificity of this cysteine protease is dependent on the catalytic residues, C25 and H163, in the S1 subsite (Methods), and therefore, a potent inhibitor should either bind or form non-covalent interactions with the catalytic residues. In this work, we consider a phytochemical to be a potential inhibitor of TMPRSS2 (respectively, cathepsin L) only if the ligand binding energy in the best docked pose is $\leq$ -8.5 kcal/mol (respectively, $\leq$ -8.0 kcal/mol) and the ligand binds to or forms non-covalent interactions with the residue D435 in TMPRSS2 (respectively, residues C25 and H163 in cathepsin L).

At the end of fourth stage, we obtained 96 phytochemicals (labelled T1-T96; Figure 4; Supplementary Table S1) as potential inhibitors of TMPRSS2 and 9 phytochemicals (labelled C1-C9; Figure 5; Supplementary Table S2) as potential inhibitors of cathepsin L. Using IMPPAT (Mohanraj et al., 2018), we provide a list of Indian medicinal plants that can produce the identified phytochemical inhibitors of TMPRSS2 and cathepsin L (Tables 1-2; Supplementary Table S3). Furthermore, we have also compiled information on potential antiviral or anti-inflammatory use in traditional medicine of the herbal sources of the identified phytochemical inhibitors of TMPRSS2 and cathepsin L (Tables 1-2; Supplementary Table S3). In Supplementary Tables S4-S5, we list the ligand binding site residues and non-covalent protein-ligand interactions for the identified phytochemical inhibitors of TMPRSS2 and cathepsin L. Finally, we have also predicted the physicochemical and ADMET properties of the identified phytochemical inhibitors of TMPRSS2 and cathepsin L (Methods; Supplementary Tables S6-S7).

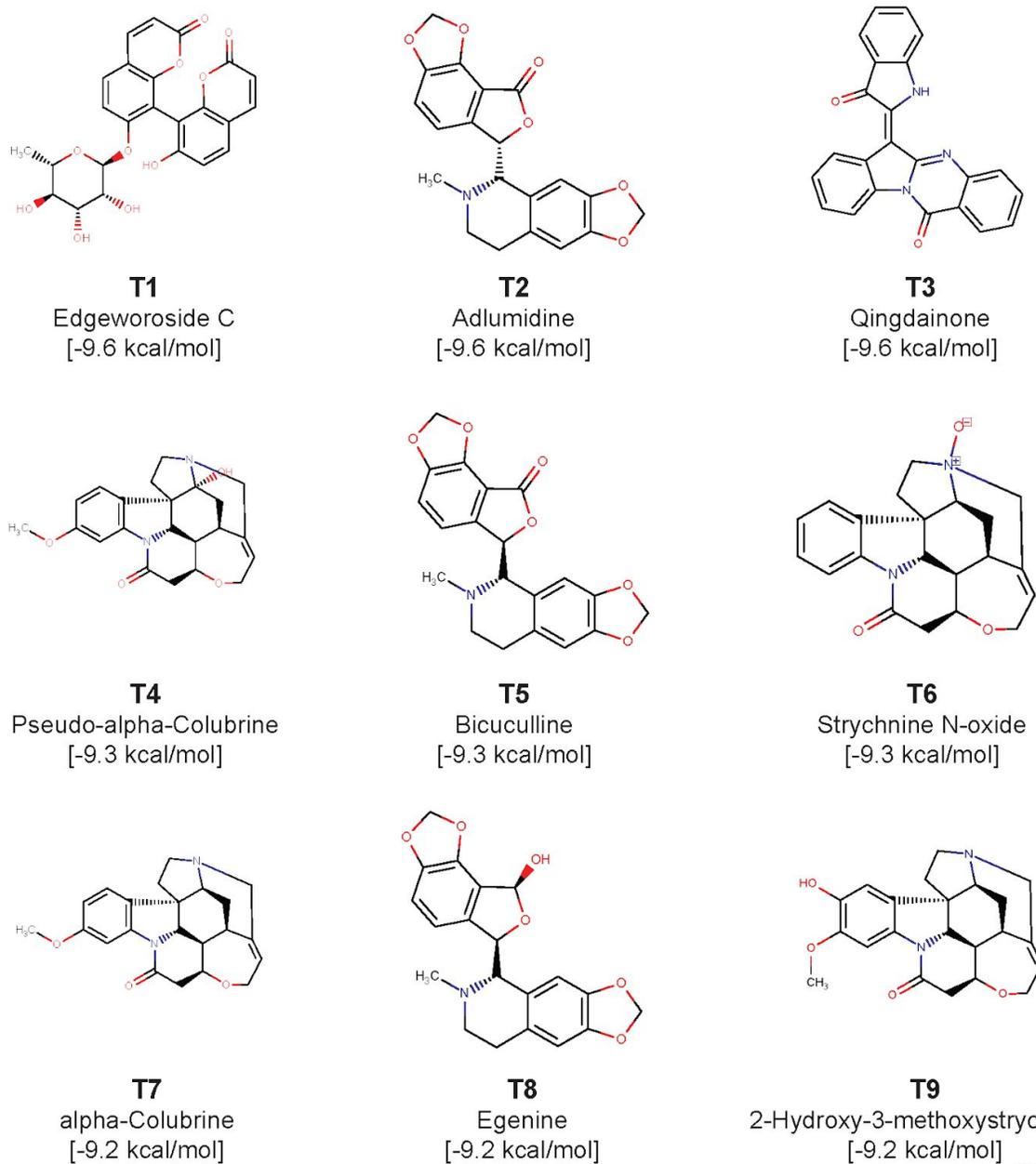

**T1**
Edgeworoside C
[-9.6 kcal/mol]

**T2**
Adlumidine
[-9.6 kcal/mol]

**T3**
Qingdainone
[-9.6 kcal/mol]

**T4**
Pseudo-alpha-Colubrine
[-9.3 kcal/mol]

**T5**
Bicuculline
[-9.3 kcal/mol]

**T6**
Strychnine N-oxide
[-9.3 kcal/mol]

**T7**
alpha-Colubrine
[-9.2 kcal/mol]

**T8**
Egenine
[-9.2 kcal/mol]

**T9**
2-Hydroxy-3-methoxystrychnine
[-9.2 kcal/mol]

**Figure 4**

**Figure 4. (Colour online)** Molecular structures of the top 9 phytochemical inhibitors
(compounds T1-T9) of TMPRSS2. For each inhibitor, the figure shows the 2D structure,
common name and docked binding energy of the ligand with TMPRSS2.

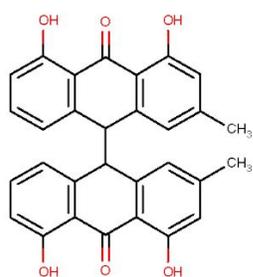

**C1**
Ararobinol
[-8.9 kcal/mol]

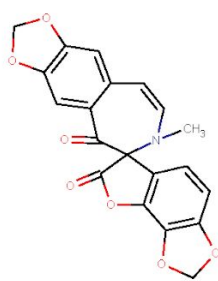

**C2**
(+)-Oxoturkiyenine
[-8.3 kcal/mol]

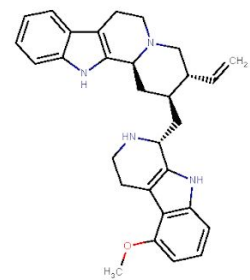

**C3**
3Alpha,17Alpha-Cinchophylline
[-8.3 kcal/mol]

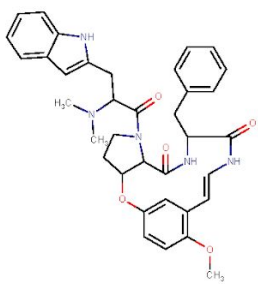

**C4**
Rugosanine B
[-8.2 kcal/mol]

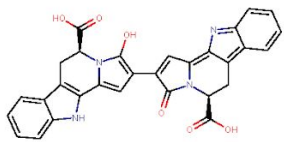

**C5**
Trichotomine
[-8.2 kcal/mol]

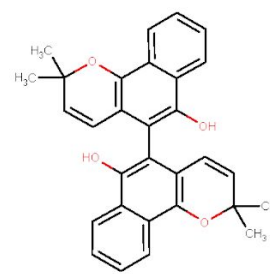

**C6**
Tectol
[-8.1 kcal/mol]

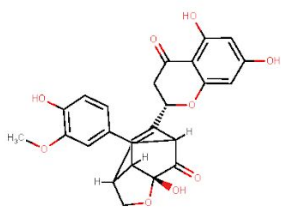

**C7**
Silymonin
[-8.1 kcal/mol]

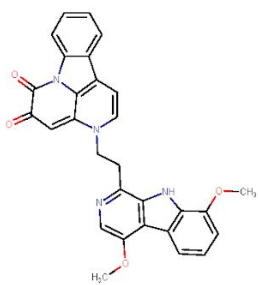

**C8**
Picrasidine M
[-8.0 kcal/mol]

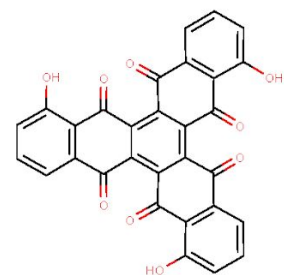

**C9**
Trisjuglone
[-8.0 kcal/mol]

**Figure 5**

**Figure 5. (Colour online)** Molecular structures of the top 9 phytochemical inhibitors (compounds C1-C9) of cathepsin L. For each inhibitor, the figure shows the 2D structure, common name and docked binding energy of the ligand with cathepsin L.

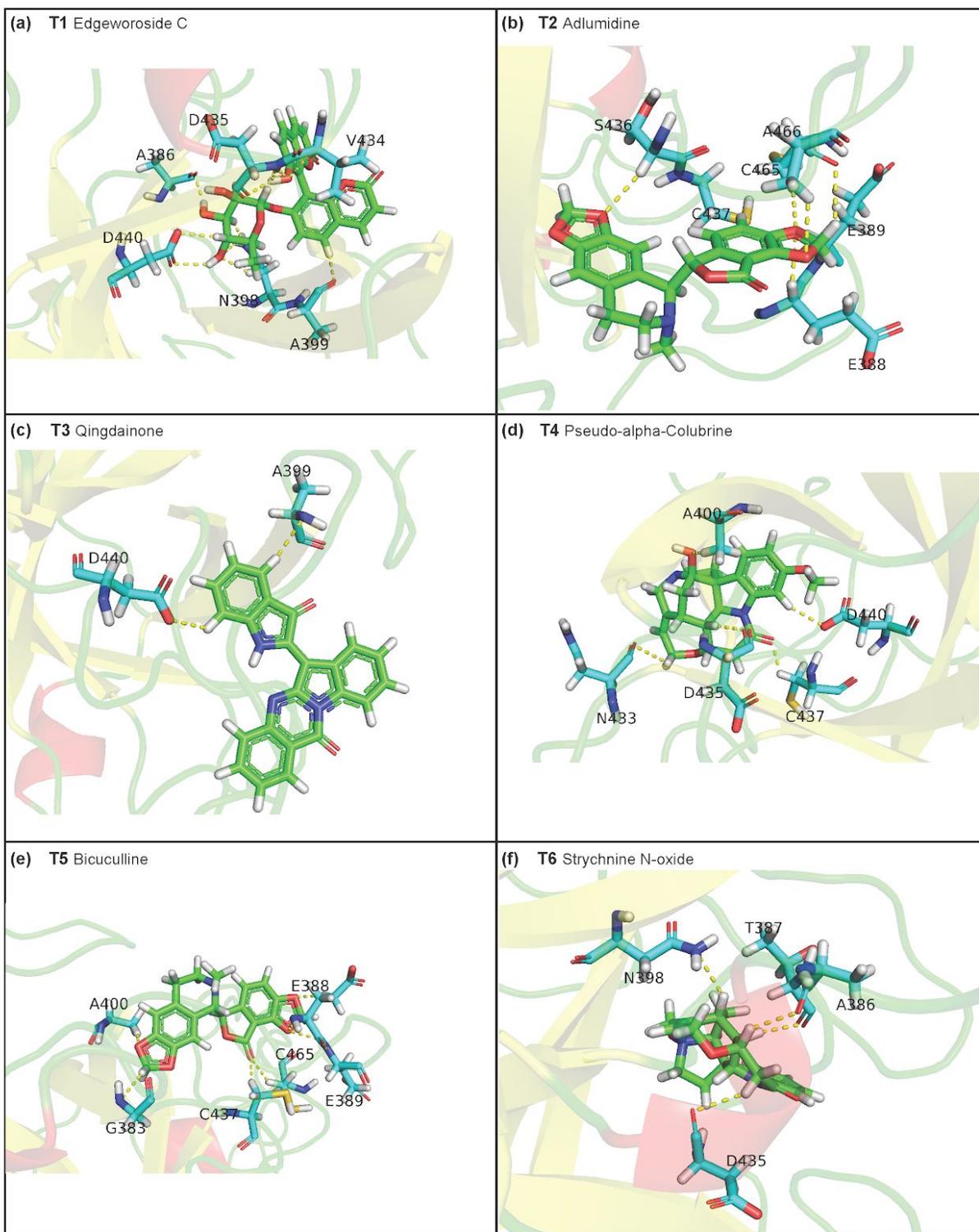

**Figure 6**

**Figure 6. (Colour online)** Cartoon representation of the protein-ligand interactions of the phytochemical inhibitors of TMPRSS2. Interactions of TMPRSS2 residues with atoms of (a) T1, (b) T2, (c) T3, (d) T4, (e) T5, and (f) T6. The carbon atoms of the ligand are shown in green colour while the carbon atoms of the amino acid residues in TMPRSS2 are shown in cyan colour. TMPRSS2 residues interacting with the ligand atoms via hydrogen bonds or $\pi$-$\pi$ stacking are labelled with the corresponding single letter residue code along with their position in the protein sequence. The hydrogen bonds and $\pi$-$\pi$ stacking are displayed using yellow and red dotted lines, respectively.

## 3.2 Potential phytochemical inhibitors of TMPRSS2

As mentioned above, we have identified 96 potential natural product inhibitors of TMPRSS2 by computational screening of 14011 phytochemicals produced by Indian medicinal plants, and these 96 compounds labelled T1-T96 are listed in Supplementary Table S1 along with their Pubchem identifier, common name, IUPAC name and structure in SMILES format. In this section, we provide a detailed discussion of the top 9 phytochemical inhibitors (labelled as T1-T9) whose binding energies in the best docked poses with TMPRSS2 are ≤ -9.2 kcal/mol. Figure 4 displays the structure of these top 9 phytochemical inhibitors of TMPRSS2 and Table 1 provides a list of Indian medicinal plants that can produce them.

Phytochemical T1, Edgeworoside C, is among three phytochemicals with binding energy of -9.6 kcal/mol. T1 is a coumarin produced by *Edgeworthia gardneri*, a medicinal plant consumed as herbal tea in Tibet (Zhao et al., 2015). In traditional medicine, *Edgeworthia gardneri* has been used to treat metabolic disorders including diabetes (Gao et al., 2015, 2016). Figure 6a shows the TMPRSS2 residues that form hydrogen bonds or $\pi$-$\pi$ stacking interactions with T1. T1 forms 12 hydrogen bonds with residues A386, N398, A399, V434, D435 and D440 of TMPRSS2. The phenolic hydroxyl group of T1 acts as both acceptor and donor forming O-H···O and N-H···O type hydrogen bonds with the substrate binding residue D435 and C-H···O type hydrogen bond with residue V434. The hydroxyl groups attached to the pyran ring of T1 form hydrogen bonds with residues A386, N398 and D440. Further, T1 forms hydrophobic contacts with residues E260, I381, A400, N433, and A466.

**Table 1:** Herbal sources of top 9 phytochemical inhibitors of TMPRSS2. For each phytochemical, the table gives the symbol, docked binding energy, common name and plant source. Plant sources which have been reported to have antiviral or anti-inflammatory use in traditional medicine literature are shown in bold and marked with an [*] sign.

| Phytochemical symbol | Binding energy (kcal/mol) | Common name | Plant source |
|---|---|---|---|
| T1 | -9.6 | Edgeworoside C | *Edgeworthia gardneri* |
| T2 | -9.6 | Adlumidine | ***Fumaria indica* [*]** |
| T3 | -9.6 | Qingdainone | ***Strobilanthes cusia* [*]** |
| T4 | -9.3 | Pseudo-alpha-Colubrine | ***Strychnos nux-vomica* [*]** |
| T5 | -9.3 | Bicuculline | ***Fumaria indica* [*], *Corydalis govaniana* [*], *Nerium oleander* [*]** |
| T6 | -9.3 | Strychnine N-oxide | ***Strychnos nux-vomica* [*], *Strychnos ignatii* [*], *Strychnos colubrina* [*]** |
| T7 | -9.2 | alpha-Colubrine | ***Strychnos nux-vomica* [*], *Strychnos ignatii* [*], *Strychnos colubrina* [*]** |
| T8 | -9.2 | Egenine | ***Fumaria vaillantii* [*]** |
| T9 | -9.2 | 2-Hydroxy-3-methoxystrychnine | ***Strychnos nux-vomica* [*]** |

Phytochemical T2, Adlumidine, also has binding energy of -9.6 kcal/mol. T2 is produced by *Fumaria indica*, a herb used in traditional medicine to treat fever, cough, skin ailments and urinary diseases (Gupta et al., 2012). Figure 6b shows the TMPRSS2 residues that form hydrogen bonds or $\pi$-$\pi$ stacking interactions with T2. The two 1,3-dioxole groups present in T2 facilitate the formation of an extensive hydrogen bond network with E388, E389, S436, C465 and A466. T2 also forms C-H⋯S type hydrogen bond with C437. Further, T2 forms hydrophobic contacts with residues E260, I381, S382, T387, N398, A399 and A400.

Phytochemical T3, Qingdainone, also has binding energy of -9.6 kcal/mol. T3 is a quinazoline alkaloid produced by *Strobilanthes cusia*, a herb with antiviral activity (Tsai et al., 2020). Figure 6c shows the TMPRSS2 residues that form hydrogen bonds or $\pi$-$\pi$ stacking interactions with T3. TMPRSS2 residue D440 forms C-H···O type hydrogen bond with T3 whereas residue A399 forms C-H···N type hydrogen bond with T3. Further, T3 forms hydrophobic contacts with residues I381, S382, T387, E388, N398, A400, D440, C465 and A466.

Phytochemicals T4 (Pseudo-alpha-Colubrine), T6 (Strychnine N-oxide), T7 (alpha-Colubrine) and T9 (2-Hydroxy-3-methoxystrychnine) have binding energies of -9.3 kcal/mol, -9.3 kcal/mol, -9.2 kcal/mol and -9.2 kcal/mol, respectively. These four phytochemicals are monoterpenoid indole alkaloids produced by *Strychnos nux-vomica*. The herb *Strychnos nux-vomica* is used in traditional Indian medicine and its alkaloids have been shown to exhibit anti-inflammatory, anti-oxidant, anti-tumor and hepatoprotective activities (Mitra et al., 2011). Note that *Strychnos nux-vomica* is a poisonous plant whose seeds are extensively used in Ayurveda only after proper detoxification procedure called *Shodhana* described in Ayurvedic texts (Mitra et al., 2011). Figure 6d shows the TMPRSS2 residues that form hydrogen bonds or $\pi$-$\pi$ stacking interactions with T4. T4 forms C-H···O type hydrogen bonds with residues A400, N433, D435 (substrate binding residue), C437 and D440. Further, T4 forms hydrophobic contacts with residues E260, I381, S382, T387, N398, A399, V434, D440 and A466. Figure 6f shows the TMPRSS2 residues that form hydrogen bonds or $\pi$-$\pi$ stacking interactions with T6. T6 forms a C-H···N type hydrogen bond with residue N398. The substrate binding residue D435 also forms a C-H···O type hydrogen bond with T6. Further, T6 forms hydrophobic contacts with residues N398, A400, V434 and A466. Supplementary Figure S3a shows the TMPRSS2 residues that form hydrogen bonds or $\pi$-$\pi$ stacking interactions with T7. T7 forms five C-H···O type hydrogen bonds with residues N433, D435 (substrate binding residue), C437 and D440. Further, T7 forms hydrophobic contacts with residues E260, T387, N398, A399, A400, V434 and A466. Supplementary Figure S3c shows the TMPRSS2 residues that form hydrogen bonds or $\pi$-$\pi$ stacking interactions with T9. The phenolic hydroxyl group of T9 forms hydrogen bonds with residues S382 and A399. The substrate binding site D435 also forms a C-H···O type hydrogen bond with T9. Further, T9 forms hydrophobic contacts with residues E260, N398, A399, A400 and V434.

Phytochemical T5, Bicuculline, has binding energy of -9.3 kcal/mol, and it is a stereoisomer of T2. T5 is an isoquinoline alkaloid and is produced by herbs *Corydalis govaniana*, *Nerium oleander* and *Fumaria indica*. Figure 6e shows the TMPRSS2 residues that form hydrogen bonds or $\pi$-$\pi$ stacking interactions with T5. The two 1,3-dioxole groups present in T5 facilitate the formation of an extensive hydrogen bond network with residues E388, E389 and A400. The Furan-2-one ring also forms a C-H···O type hydrogen bond with C437. Further, T5 forms hydrophobic contacts with residues E260, T387, E388, N398, A399 and A466.

Phytochemical T8, Egenine, has binding energy of -9.2 kcal/mol. T8 is an isoquinoline alkaloid produced by *Fumaria vaillantii*. In traditional medicine, *Fumaria vaillantii* has been reported to exhibit antifungal, anti-inflammatory and anti-psychotic activities (Khare, 2007). Supplementary Figure S3b shows the TMPRSS2 residues that form hydrogen bonds or $\pi$-$\pi$ stacking interactions with T8. The two 1,3-dioxole groups present in T8 form hydrogen bonds with G383, E388, E389 and A400. One of the hydroxyl groups present in T8 forms C-H···O type hydrogen bond with residue C437. Further, T8 form hydrophobic contacts with residues T387, A399, E388, N398, E260, and A466.

### 3.3 Potential inhibitors of Cathepsin L

As mentioned above, we have identified 9 potential natural product inhibitors of cathepsin L by computational screening of 14011 phytochemicals produced by Indian medicinal plants, and these compounds labelled C1-C9 are listed in Supplementary Table S2 along with their Pubchem identifier, common name, IUPAC name and structure in SMILES format. Figure 5 displays the structure of these top 9 phytochemical inhibitors of cathepsin L and Table 2 provides a list of Indian medicinal plants that can produce them.

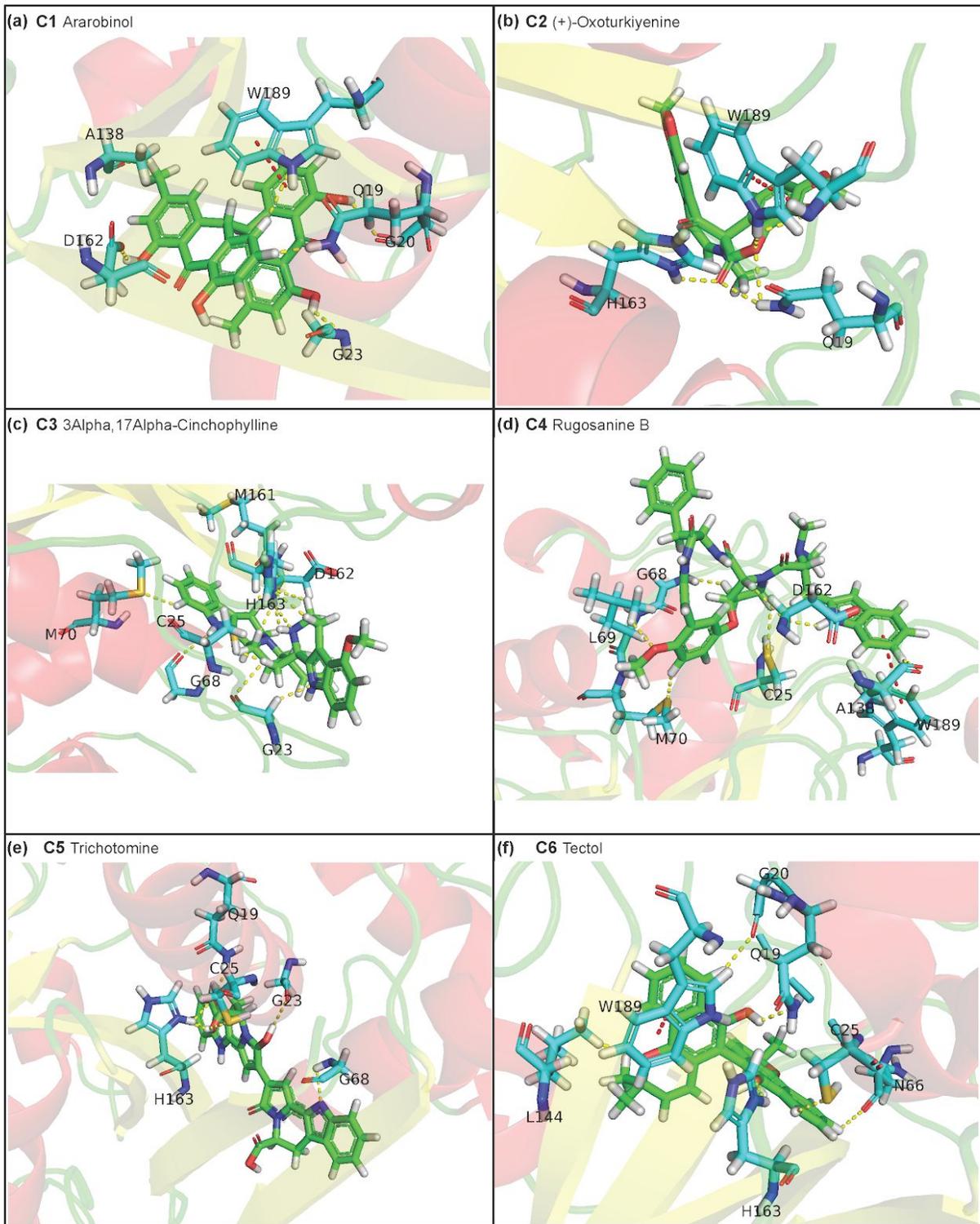

**Figure 7**

**Figure 7. (Colour online)** Cartoon representation of the protein-ligand interactions of the phytochemical inhibitors of cathepsin L. Interactions of cathepsin L residues with atoms of (a) C1, (b) C2, (c) C3, (d) C4, (e) C5, and (f) C6. The carbon atoms of the ligand are shown in green colour while the carbon atoms of the amino acid residues in cathepsin L are shown in cyan colour. Cathepsin L residues interacting with the ligand atoms via hydrogen bonds or $\pi$-$\pi$ stacking are labelled with the corresponding single letter residue code along with their position in the protein sequence. The hydrogen bonds and $\pi$-$\pi$ stacking are displayed using yellow and red dotted lines, respectively.

**Table 2:** Herbal sources of top 9 phytochemical inhibitors of Cathepsin L. For each phytochemical, the table gives the symbol, docked binding energy, common name and plant source. Plant sources which have been reported to have antiviral or anti-inflammatory use in traditional medicine literature are shown in bold and marked with an [*] sign.

| Phytochemical symbol | Binding energy (kcal/mol) | Common name | Plant source |
|---|---|---|---|
| C1 | -8.9 | Ararobinol | *Senna occidentalis* [*] |
| C2 | -8.3 | (+)-Oxoturkiyenine | *Hypecoum pendulum* |
| C3 | -8.3 | 3Alpha,17Alpha-Cinchophylline | *Cinchona calisaya* [*] |
| C4 | -8.2 | Rugosanine B | *Ziziphus rugosa* [*] |
| C5 | -8.2 | Trichotomine | *Clerodendrum trichotomum* [*] |
| C6 | -8.1 | Tectol | *Tectona grandis* [*], *Tecomella undulata* [*] |
| C7 | -8.1 | Silymonin | *Silybum marianum* [*] |
| C8 | -8 | Picrasidine M | *Picrasma quassioides* [*] |
| C9 | -8 | Trisjuglone | *Juglans regia* [*] |

Phytochemical C1, Ararobinol, has binding energy of -8.9 kcal/mol. C1 is a bianthraquinone produced by herb *Senna occidentalis* used in Ayurveda. In traditional medicine, *Senna occidentalis* has been reported for antibacterial, antifungal, anti-inflammatory, anti-diabetic and anti-cancer activities (J. P. Yadav et al., 2010). Interestingly, there is a Chinese patent application (ZHU SHOUHUI YANG, 2005) on potential use of C1 to treat human influenza virus infections. Figure 7a shows the cathepsin L residues that form hydrogen bonds or $\pi$-$\pi$ stacking interactions with C1. The phenolic hydroxyl groups of C1 form hydrogen bonds with residues G20 and G23. Residues Q19, A138 and D162 form C-H···N, C-H···O and O-H···O type of hydrogen bonds, respectively, with C1. Also, the residue W189 forms both face-to-edge and face-to-face type of $\pi$-$\pi$ stacking interaction with C1. Further, C1 forms hydrophobic contacts with residues C25, G139, L144, H163 and W189.

Phytochemical C2, (+)-Oxoturkiyenine, has binding energy of -8.3 kcal/mol. C2 is an isoquinoline-derived alkaloid produced by the herb *Hypecoum pendulum* (Mete & Gözler, 1988). Figure 7b shows the cathepsin L residues that form hydrogen bonds or $\pi$-$\pi$ stacking interactions with C2. The 2,5-dihydro-furan ring present in C2 forms two N-H···O type hydrogen bonds with residue Q19. Also, one of the 1,3-dioxole groups in C2 forms N-H···O type hydrogen bond with residue W189. The catalytic residue H163 forms N-H···O type hydrogen bond with C2. Also, the residue W189 forms a face-to-edge type of $\pi$-$\pi$ stacking interaction with C2. Further, C2 forms hydrophobic contacts with residues G139, H140, H163 and W189.

Phytochemical C3, 3Alpha,17Alpha-Cinchophylline, has binding energy of -8.3 kcal/mol. C3 is a cinchophylline-type of alkaloid produced by the herb *Cinchona calisaya*. The *Cinchona* alkaloids have been reported for their antimicrobial, antiparasitic and anti-inflammatory activities (Gurung & De, 2017). Figure 7c shows the cathepsin L residues that form hydrogen bonds or $\pi$-$\pi$ stacking interactions with C3. C3 forms 9 hydrogen bonds with residues of cathepsin L. One of the catalytic residue C25 forms C-H···S type hydrogen bond with C3. The other catalytic residue H163 forms C-H···N and N-H···N type hydrogen bonds with C3. Further, the two pyrrole rings present in C3 form hydrogen bonds with residues G23 and G68. Lastly, M70 forms a C-H···S type hydrogen bond with C3. Further, C3 forms hydrophobic contacts with residues Q21, C22, L69, M70, A135 and W189.

Phytochemical C4, Rugosanine B, has binding energy of -8.2 kcal/mol. C4 is a cyclopeptide alkaloid produced by the bark of *Ziziphus rugosa* (Tripathi et al., 1989). Various parts of *Ziziphus rugosa* have been reported for their antibacterial, antifungal, anti-inflammatory and analgesic activities (*The Wealth of India*, 2000e; A. Yadav & Singh, 2010). Figure 7d shows the cathepsin L residues that form hydrogen bonds or $\pi$-$\pi$ stacking interactions with C4. The pyrrole ring present in C4 forms a N-H···O type hydrogen bond with residue D162. The catalytic residue C25 forms a C-H···S type hydrogen bond with the pyrrolidine group of C4. Also, the residue W189 forms a face-to-edge type of $\pi$-$\pi$ stacking interaction with C4. Further, C4 forms hydrophobic contacts with residues G23, A135, A138, D162, H163, G164, W189 and A214.

Phytochemical C5, Trichotomine, has binding energy of -8.2 kcal/mol. C5 is a bisindole alkaloid present in *Clerodendrum trichotomum* (*Medicinal and Aromatic Plants IX.*, 2011). *Clerodendrum trichotomum* has been reported for its use to treat cold, hypertension, rheumatism, dysentery and other inflammatory diseases (J.-H. Wang et al., 2018). Figure 7e shows the cathepsin L residues that form hydrogen bonds or $\pi$-$\pi$ stacking interactions with C5. The carboxylic acid group present in C5 forms hydrogen bonds with residues Q19, C25 and H163. The indole ring of C5 forms a hydrogen bond with residue G68. Further, C5 forms hydrophobic contacts with residues G23, G67, G68, L69 and Y72.

Phytochemical C6, Tectol, has binding energy of -8.1 kcal/mol. C6 is a naphthoquinone derivative (Sumthong et al., 2008) present in *Tectona grandis* and *Tecomella undulata*. *Tectona grandis* has been reported to have anti-inflammatory and antiparasitic activities (Khare, 2007). *Tecomella undulata* has been used to treat syphilis and also reported to have anti-inflammatory and anti-HIV activities (Jain et al., 2012). Figure 7f shows the cathepsin L residues that form hydrogen bonds or $\pi$-$\pi$ stacking interactions with C6. The phenolic hydroxyl group of C6 forms a O-H···N type hydrogen bond with residue Q19. The pyran group of C6 is involved in a C-H···O type hydrogen bond with residue L144. The catalytic residue C25 forms a C-H···S type hydrogen bond with C6. The other catalytic residue H163 forms a N-H···O type hydrogen bond with C6. Also, the residue W189 forms both face-to-face and face-to-edge type of $\pi$-$\pi$ stacking interaction with C6. Further, C6 forms hydrophobic contacts with G23, L144 and W189.

Phytochemical C7, Silymonin, has binding energy of -8.1 kcal/mol. C7 is a flavanolignan (Bosisio et al., 1992) present in *Silybum marianum*. *Silybum marianum* has been used as a hepatoprotective agent and is reported to have anti-oxidant and anti-inflammatory activities (Bahmani et al., 2015). Supplementary Figure S4a shows the cathepsin L residues that form hydrogen bonds or $\pi$-$\pi$ stacking interactions with C7. C7 has four hydroxyl groups which help in the formation of an extensive network of hydrogen bonds with residues Q21, N66, G139, D162 and H163. Also, the residue W189 forms a face-to-edge type of $\pi$-$\pi$ stacking interaction with C7. Further, C7 forms hydrophobic contacts with residues G23, A138, L144, H163 and W189.

Phytochemical C8, Picrasidine M, has binding energy of -8.0 kcal/mol. C8 is a β-carboline alkaloid present in *Picrasma quassioides*. *Picrasma quassioides* has been reported to have antiviral and antifungal activities (*The Wealth of India*, 2000d). Supplementary Figure S3b shows the cathepsin L residues that form hydrogen bonds or $\pi$-$\pi$ stacking interactions with C8. The carboxylic group of residue D162 forms two C-H⋯O type hydrogen bonds with C8. Also, residues M70 and G23 form hydrogen bonds of type C-H⋯S and C-H⋯O, respectively, with C8. Also, the residue W189 forms a face-to-edge type of $\pi$-$\pi$ stacking interaction with C8. Further, C8 forms hydrophobic contacts with residues G23, L69, D162 and W189.

Phytochemical C9, Trisjuglone, has binding energy of -8.0 kcal/mol. C9 is a naphthoquinone present in *Juglans regia* (i.e., walnut). In traditional medicine, *Juglans regia* has been reported to have anti-inflammatory, antifungal and antimicrobial activities (Khare, 2007). Supplementary Figure S3c shows the cathepsin L residues that form hydrogen bonds or $\pi$-$\pi$ stacking interactions with C9. The benzoquinone moiety present in C9 forms two C-H⋯O type hydrogen bonds with residues Q21 and G23. In contrast, the other catalytic residue H163 forms a N-H⋯O type hydrogen bond with C9. Also, the residue W189 forms a face-to-edge type of $\pi$-$\pi$ stacking interaction with C9. Further, C9 forms hydrophobic contacts with residues Q21, G23, C25, L144 and W189.

## 4. Conclusion

Current COVID-19 pandemic is a serious threat to humankind. As of 30 May 2019, COVID-19 has led to more than 3,65,000 deaths worldwide. Due to the absence of approved therapeutics or vaccines against SARS-CoV-2, several countries have been forced to implement partial or complete lockdown measures to restrict infection spread, however, such measures have in turn resulted in an economic catastrophe. Consequently, there is an urgent need to develop antivirals and vaccines against SARS-CoV-2 to protect humankind. In this direction, protein-ligand docking is a powerful computational method to expedite the search for anti-COVID drugs by rapid identification of promising lead molecules. Here, we have used molecular docking in the search of natural compound inhibitors of two human proteases, TMPRSS2 and cathepsin L, that are key host factors in SARS-CoV-2 infection (Hoffmann, Kleine-Weber, et al., 2020; Ou et al., 2020; Shang, Wan, et al., 2020).

Since early civilization, humans have used medicinal plants in different systems of traditional medicine to treat various ailments (Yuan et al., 2016). Specifically, traditional systems of Indian medicine including Ayurveda, Siddha and Unani have over centuries acquired invaluable knowledge on medicinal plants spanning the rich biodiversity of the subcontinent for treating various ailments including viral infections (Mohanraj et al., 2018). As plant-based natural products have been an indomitable source of lead molecules in the course of modern drug discovery (Newman & Cragg, 2012), it is worthwhile to search for potential anti-COVID drugs among phytochemicals of Indian medicinal plants. In this direction, some of us have built IMPPAT (Mohanraj et al., 2018), the largest resource on phytochemicals of Indian medicinal plants to facilitate natural product based drug discovery. In this work, we have performed virtual screening of 14010 phytochemicals that are produced by Indian medicinal plants to identify potential inhibitors of key host factors, TMPRSS2 and cathepsin L, for SARS-CoV-2 infection.

We have predicted 96 potential phytochemical inhibitors of TMPRSS2, of which the top candidates with least binding energy of -9.6 kcal/mol are Edgeworoside C, Adlumidine and Qingdainone (Figure 4). We have also predicted 9 potential phytochemical inhibitors of cathepsin L, of which the top candidate with least binding energy of -8.9 kcal/mol is Ararobinol (Figure 5). Propitiously, Ararobinol is mentioned in a Chinese patent application (ZHU SHOUHUI YANG, 2005) for its potential to treat human influenza virus infections. Furthermore, most of the herbal sources of the identified phytochemical inhibitors of TMPRSS2 and cathepsin L are reported to have antiviral or anti-inflammatory use in traditional medicine (Tables 1-2). Additional *in vitro* and *in vivo* testing of the identified phytochemical inhibitors of TMPRSS2 and cathepsin L is needed before these molecules can enter clinical trials against COVID-19. In conclusion, we expect the natural product inhibitors identified in this computational study for TMPRSS2 and cathepsin L will likely inform future research toward natural product-based anti-COVID therapeutics.

**Abbreviations**

COVID-19          Coronavirus disease 2019

SARS-CoV-2        Severe acute respiratory syndrome coronavirus 2

SARS-CoV          Severe acute respiratory syndrome coronavirus

MERS-CoV          Middle east respiratory syndrome coronavirus

TMPRSS2           Transmembrane Protease Serine 2

RBD               Receptor binding domain

ACE-2             Angiotensin converting enzyme 2

ORF               Open reading frame

FDA               Food and drug administration

ADMET             Absorption, distribution, metabolism, excretion and toxicity

**Acknowledgements**

We thank S. Krishnaswamy, Dhiraj Kumar, Vinay Nandicoori and Pinaki Saha for discussions. AS and RPV thank the computational staff of IMSc for maintaining resources during the Covid-19 associated lockdown. AS acknowledges financial support from the Science and Engineering Research Board (SERB) India through the award of a Ramanujan fellowship (SB/S2/RJN-006/2014), Department of Atomic Energy (DAE) India, Max Planck Society

Germany through the award of a Max Planck Partner Group in Mathematical Biology and The Abdus Salam International Centre for Theoretical Physics Italy through the award of a Simons Associateship. HSB and AR acknowledge the computational facilities of NISER. The funders have no role in study design, data collection, data analysis, manuscript preparation or decision to publish.

**Author contributions**

AS conceived the study. RPV, NR and AS compiled datasets. RPV and AS performed the virtual screening including homology modeling and molecular docking. RPV, AR and HSB performed the prediction and analysis of non-covalent interactions between ligand and proteins. RPV, HSB and AS analyzed the results. RPV, AR, HSB and AS wrote the manuscript. All authors have read and approved the manuscript.

**Declaration of Competing Interests**

The authors declare that they have no competing interests.

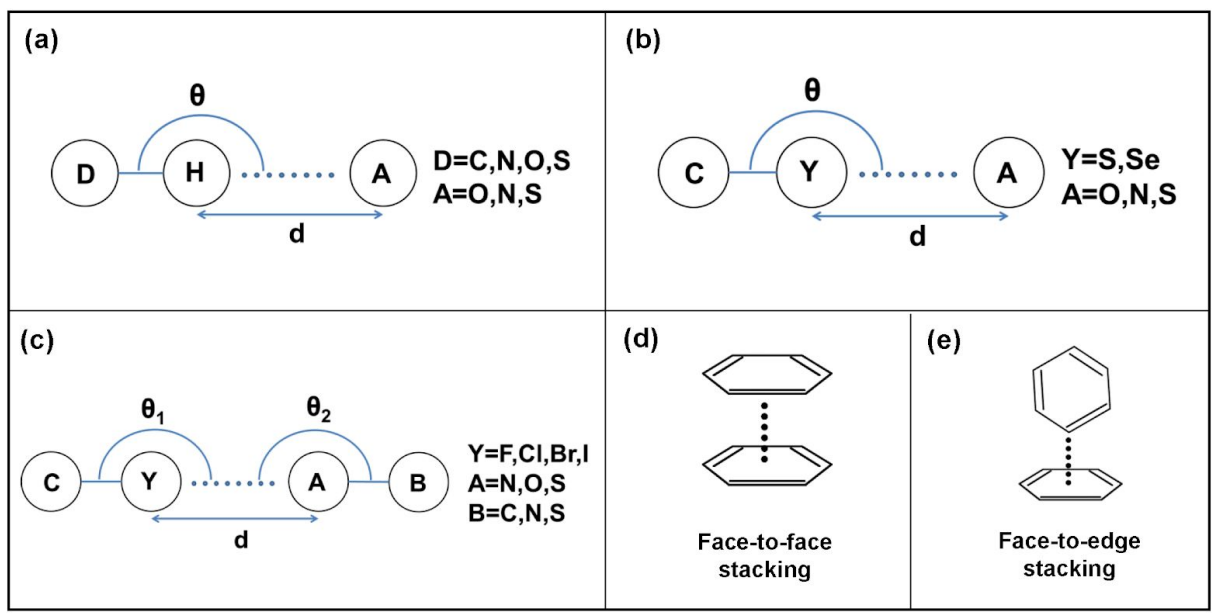

**Figure S1.** Geometric criteria for the identification of protein-ligand interactions. (a) Hydrogen bond, (b) Chalcogen bond, (c) Halogen bond, (d) face-to-face $\pi$-$\pi$ stacking, and (e) face-to-edge $\pi$-$\pi$ stacking.

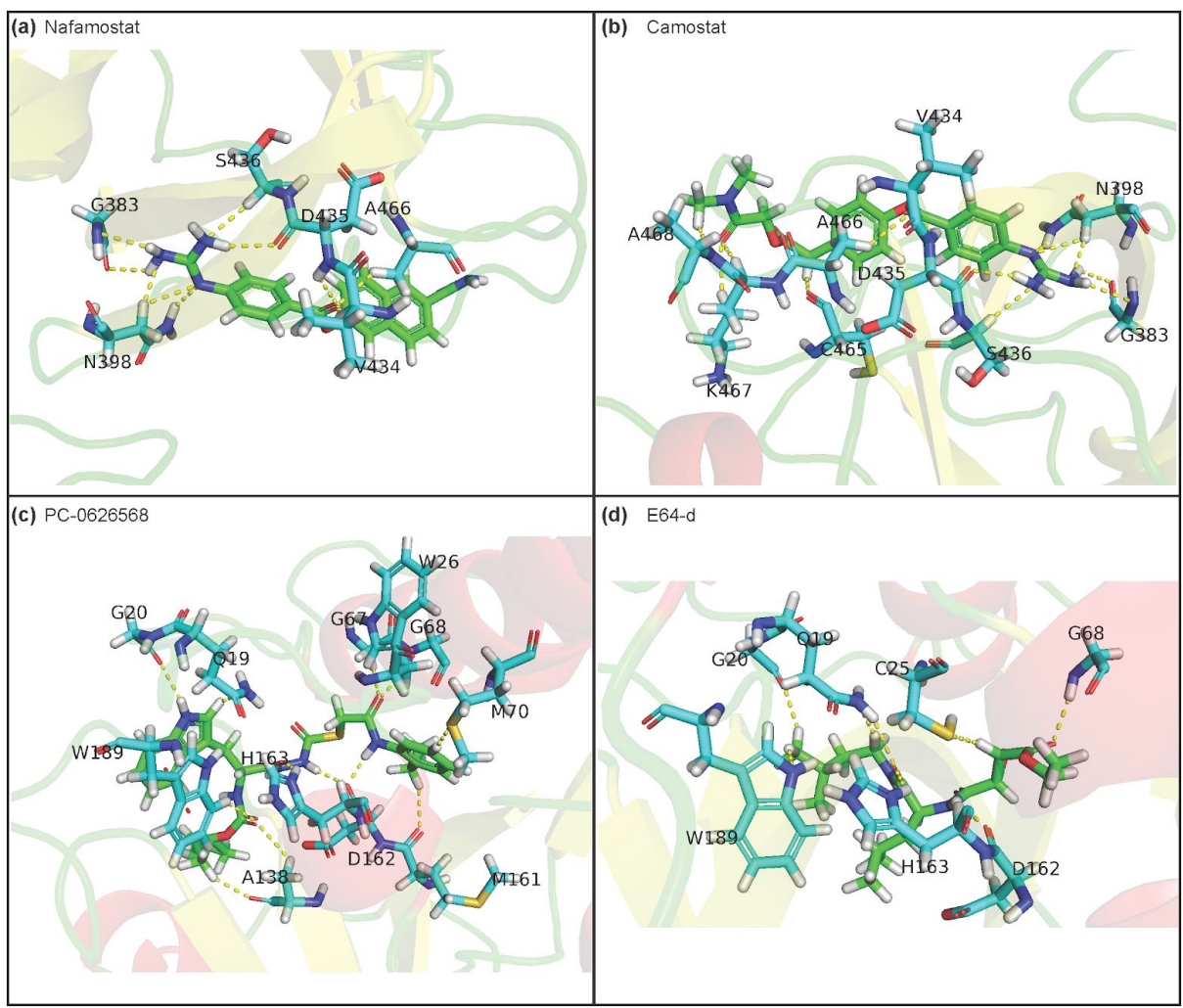

**Figure S2.** Cartoon representation of the protein-ligand interactions of the known inhibitors of TMPRSS2 and cathepsin L. Interactions of TMPRSS2 residues with atoms of (a) Nafamostat, and (b) Camostat. Interactions of cathepsin L residues with atoms of (c) PC-0626568, and (d) E-64d. The carbon atoms of the ligand are shown in green colour while the carbon atoms of the amino acid residues in TMPRSS2 or cathepsin L are shown in cyan colour. TMPRSS2 or cathepsin L residues interacting with the ligand atoms via hydrogen bonds or $\pi$-$\pi$ stacking are labelled with the corresponding single letter residue code along with their position in the protein sequence. The hydrogen bonds and $\pi$-$\pi$ stacking are displayed using yellow and red dotted lines, respectively.

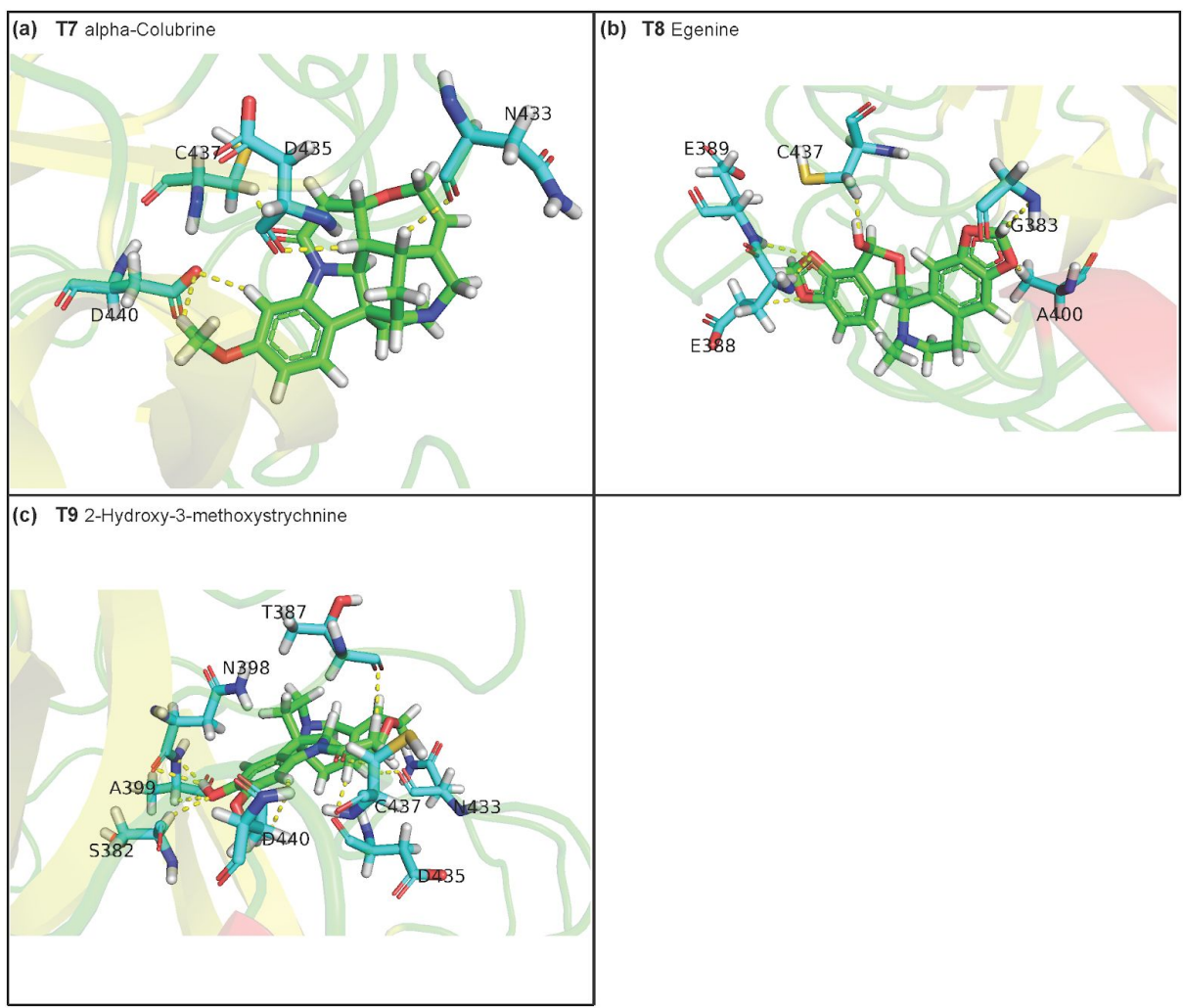

**Figure S3.** Cartoon representation of the protein-ligand interactions of the phytochemical inhibitors of TMPRSS2. Interactions of TMPRSS2 residues with atoms of (a) T7, (b) T8, and (c) T9. The carbon atoms of the ligand are shown in green colour while the carbon atoms of the amino acid residues in TMPRSS2 are shown in cyan colour. TMPRSS2 residues interacting with the ligand atoms via hydrogen bonds or $\pi$-$\pi$ stacking are labelled with the corresponding single letter residue code along with their position in the protein sequence. The hydrogen bonds and $\pi$-$\pi$ stacking are displayed using yellow and red dotted lines, respectively.

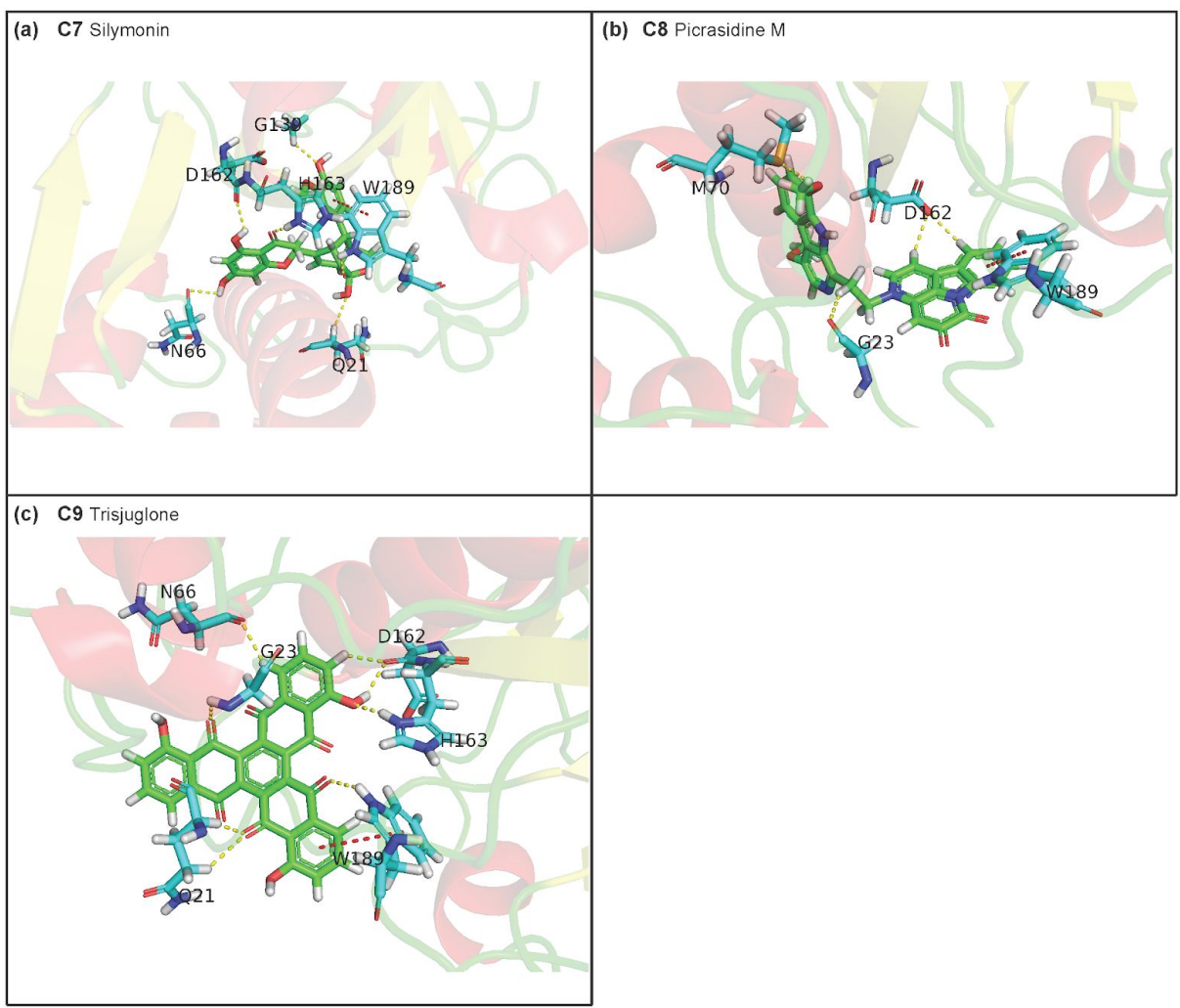

**Figure S4.** Cartoon representation of the protein-ligand interactions of the phytochemical inhibitors of cathepsin L. Interactions of cathepsin L residues with atoms of (a) C7, (b) C8, and (c) C9. The carbon atoms of the ligand are shown in green colour while the carbon atoms of the amino acid residues in cathepsin L are shown in cyan colour. Cathepsin L residues interacting with the ligand atoms via hydrogen bonds or $\pi$-$\pi$ stacking are labelled with the corresponding single letter residue code along with their position in the protein sequence. The hydrogen bonds and $\pi$-$\pi$ stacking are displayed using yellow and red dotted lines, respectively.

**Supplementary Tables**

**Table S1.** Top list of 96 phytochemical inhibitors of TMPRSS2. For each phytochemical, the table gives the symbol, docked binding energy, Pubchem identifier, common name, IUPAC name and SMILES.

**Table S2.** Top list of 9 phytochemical inhibitors of cathepsin L. For each phytochemical, the table gives the symbol, docked binding energy, Pubchem identifier, common name, IUPAC name and SMILES.

**Table S3.** Herbal sources of top 96 phytochemical inhibitors of TMPRSS2. For each phytochemical, the table gives the symbol, docked binding energy, common name and plant source. Plant sources which have been reported to have anti-viral or anti-inflammatory therapeutic use in traditional medicine literature are shown in bold and marked with an [*] sign.

**Table S4.** The table lists the ligand binding site residues and non-covalent protein-ligand interactions namely, Hydrogen bond interactions, π-π stacking interactions of face-to-face type and face-to-edge type and Hydrophobic interactions that were identified from the docked protein-ligand complexes of the top 96 phytochemical inhibitors of TMPRSS2.

**Table S5.** The table lists the ligand binding site residues and non-covalent protein-ligand interactions namely, Hydrogen bond interactions, π-π stacking interactions of face-to-face type and face-to-edge type and Hydrophobic interactions that were identified from the docked protein-ligand complexes of the top 9 phytochemical inhibitors of cathepsin L.

**Table S6.** ADMET properties of the top list of 96 potential phytochemical inhibitors of TMPRSS2. For each phytochemical, the table gives the symbol, Pubchem identifier; **several physicochemical properties** such as Molecular weight in g/mol, LogP (Partition coefficient), MolarRefractivity, TPSA (Topological Surface Area) in Å², Number of hydrogen bond acceptors, Number of hydrogen bond donors, Total number of atoms, Number of rotatable bonds, Shape complexity (fraction of carbon atoms that are sp3 hybridized), Stereochemical complexity (fraction of carbon atoms which are stereogenic); **several drug-likeness properties** such as Lipinski RO5 filter (Lipinski's Rule of 5 filter), QEDw (weighted quantitative estimate of drug-likeness), QEDuw (unweighted quantitative estimate of drug-likeness), Leadlikeness - number of violations predicted by SwissADME, Synthetic accessibility predicted by SwissADME; **several absorption and distribution properties** such as Solubility class predicted by SwissADME using ESOL method, Solubility class predicted by SwissADME using method by Ali et al, Solubility class predicted by SwissADME using Silicos-IT, Gasterointestinal absorption predicted by SwissADME, Skin permeation predicted by SwissADME (log Kp cm/s), Blood Brain Barrier permeation predicted by SwissADME, Blood Brain Barrier permeation predicted by vNNADMET, Bioavailability Score predicted by SwissADME; **several metabolism properties** such as P-glycoprotein substrate predicted by SwissADME, P-glycoprotein Substrate predicted by vNNADMET, P-glycoprotein Inhibitor predicted by vNNADMET, Cytochrome P450 1A2 inhibitor predicted by SwissADME, Cytochrome P450 1A2 Inhibitor predicted by vNNADMET, Cytochrome P450 2C19 inhibitor predicted by SwissADME, Cytochrome P450 2C19 Inhibitor predicted by vNNADMET, Cytochrome P450 2C9 inhibitor predicted by SwissADME, Cytochrome P450 2C9 Inhibitor predicted by vNNADMET, Cytochrome P450 2D6 inhibitor predicted by SwissADME, Cytochrome P450 2D6 Inhibitor predicted by vNNADMET, Cytochrome P450 3A4 inhibitor predicted by SwissADME, Cytochrome P450 3A4 Inhibitor predicted by vNNADMET, Human liver microsomal stability predicted by vNNADMET (HLM); **several toxicity properties** such as PAINS - Number of alerts predicted by SwissADME, Brenk - Number of alerts predicted by SwissADME, Drug-induced liver injury predicted by vNNADMET (DILI), Cytotoxicity predicted by vNNADMET, hERG Blocker predicted by vNNADMET, Mitochondrial toxicity predicted by vNNADMET (MMP) and Mutagenecity predicted by vNNADMET (AMES).

**Table S7.** ADMET properties of the top list of 9 potential phytochemical inhibitors of cathepsin L. For each phytochemical, the table gives the symbol, Pubchem identifier; **several physicochemical properties** such as Molecular weight in g/mol, LogP (Partition coefficient), MolarRefractivity, TPSA (Topological Surface Area) in Å², Number of hydrogen bond acceptors, Number of hydrogen bond donors, Total number of atoms, Number of rotatable bonds, Shape complexity (fraction of carbon atoms that are sp3 hybridized), Stereochemical complexity (fraction of carbon atoms which are stereogenic); **several drug-likeness properties**

such as Lipinski RO5 filter (Lipinski's Rule of 5 filter), QEDw (weighted quantitative estimate of drug-likeness), QEDuw (unweighted quantitative estimate of drug-likeness), Leadlikeness - number of violations predicted by SwissADME, Synthetic accessibility predicted by SwissADME; **several absorption and distribution properties** such as Solubility class predicted by SwissADME using ESOL method, Solubility class predicted by SwissADME using method by Ali et al, Solubility class predicted by SwissADME using Silicos-IT, Gasterointestinal absorption predicted by SwissADME, Skin permeation predicted by SwissADME (log Kp cm/s), Blood Brain Barrier permeation predicted by SwissADME, Blood Brain Barrier permeation predicted by vNNADMET, Bioavailability Score predicted by SwissADME; **several metabolism properties** such as P-glycoprotein substrate predicted by SwissADME, P-glycoprotein Substrate predicted by vNNADMET, P-glycoprotein Inhibitor predicted by vNNADMET, Cytochrome P450 1A2 inhibitor predicted by SwissADME, Cytochrome P450 1A2 Inhibitor predicted by vNNADMET, Cytochrome P450 2C19 inhibitor predicted by SwissADME, Cytochrome P450 2C19 Inhibitor predicted by vNNADMET, Cytochrome P450 2C9 inhibitor predicted by SwissADME, Cytochrome P450 2C9 Inhibitor predicted by vNNADMET, Cytochrome P450 2D6 inhibitor predicted by SwissADME, Cytochrome P450 2D6 Inhibitor predicted by vNNADMET, Cytochrome P450 3A4 inhibitor predicted by SwissADME, Cytochrome P450 3A4 Inhibitor predicted by vNNADMET, Human liver microsomal stability predicted by vNNADMET (HLM); **several toxicity properties** such as PAINS - Number of alerts predicted by SwissADME, Brenk - Number of alerts predicted by SwissADME, Drug-induced liver injury predicted by vNNADMET (DILI), Cytotoxicity predicted by vNNADMET, hERG Blocker predicted by vNNADMET, Mitochondrial toxicity predicted by vNNADMET (MMP) and Mutagenecity predicted by vNNADMET (AMES).

*and bird flu virus H5N1*. http://europepmc.org/patents/PAT/CN1985898